# Big Data Analytics-Enabled Dynamic Capabilities and Market Performance: Examining the Roles of Marketing Ambidexterity and Competitor Pressure

**Gulfam Haider[1], Laiba Zubair[1], Aman Saleem[1]**

[1]FAST School of Management, National University of Computer and Emerging Sciences, Chiniot-Faisalabad Campus, Chiniot 35400, Pakistan.

## Abstract

This study, rooted in dynamic capability theory and the developing era of Big Data Analytics, explores the transformative effect of BDA-EDCs on marketing. Ambidexterity and firm's market performance in the textile sector of Pakistan's cities. Specifically, focusing on the firms who directly deal with customers, investigates the nuanced role of BDA-EDCs in textile retail firms' potential to navigate market dynamics. Emphasizing the exploitation component of marketing ambidexterity, the study investigated the mediating function of marketing ambidexterity and the moderating influence of competitive pressure. Using a survey questionnaire, the study targets key choice-makers in textile firms of Faisalabad, Chiniot and Lahore, Pakistan. The PLS-SEM model was employed as an analytical technique, allows for a full examination of the complicated relations between BDA-EDCs, marketing ambidexterity, rival pressure, and market performance. The study Predicting a positive impact of Big Data on marketing ambidexterity, with a specific emphasis on exploitation. The study expects this exploitation-orientated marketing ambidexterity to significantly enhance the firm's market performance. This research contributes to the existing literature on dynamic capabilities-based frameworks from the perspective of the retail segment of textile industry. The study emphasizes the role of BDA-EDCs in the retail sector, imparting insights into the direct and indirect results of BDA-EDCs on market performance inside the retail area. The study's novelty lies in its contextualization of BDA-EDCs in the textile zone of Faisalabad, Lahore and Chiniot, providing a unique perspective on the effect of BDA on marketing ambidexterity and market performance in firms. Methodologically, the study uses numerous samples of retail sectors to make sure broader universality, contributing realistic insights.

## INTRODUCTION

Businesses these days employ enormous amounts of data. companies generate massive amounts of transactional data, capturing millions of data approximately their customers, suppliers, and operations Manyika et al., (2011). These massive amounts of data are known as Big Data. Today, companies heavily rely on mobile and internet to acquire competitive gain Mikalef et al., (2019b). It is another point to believe that this era is the era of data Mikalef et al., (2019b). Big Data refers to massive and different datasets that contain several types and volumes of data Brown et al., (2011); Rialti et al., (2019). This study investigates how modern tools consisting of Big Data Analytics can assist businesses in adapting to these tools and enhancing their plans. BDA possesses the ability to transform traditional business practices essentially Rialti et al., (2019). Our study focuses on how those tools affect an organization's potential to efficiently optimize and leverage its resources. Recent studies have encouraged more research on the connection between BDA and organization performance through incorporating an element of data analytics and multiple viewpoints into the studies Dubey et al., (2022); Zhang et al., (2022).

This research take place in the textile sector of Faisalabad, Chiniot, Lahore, Pakistan and tries to discover how those techniques influence a company's market performance in the face of stiff competition. Managers nowadays may now gain a more understanding of their businesses, competitors, and clients than ever before because of data Rialti et al., (2019). Compared to their competitors, online businesses that utilize BDA into their businesses noticed a five to six percentage higher increase in returns

Akter & Wamba, (2016); Gaur et al., (2021). Research is still being done on how BDAEDCs, market performance, and marketing ambidexterity are associated. This research aims to investigate the characters of marketing ambidexterity and competitor pressure in firms' performance, aiming to fill this gap in understanding.

The massive amount of data in today's corporate landscape offers great chances for better decision-making, but many firms struggle to recognize their full potential. These studies delved into how Big Data Analytics (BDA) can be a valuable tool in helping companies in navigating this difficult balance. Big Data Analytics (BDA) could be a wide-ranging approach to overseeing, preparing, and analyzing enormous sums of information in terms of five Vs, the objective is to deliver critical experiences that include to performance evaluation, and competitive points of interest (Fosso Wamba et al., (2015); Mohamed et al., (2020); Kalyar et al., (2023) Ahmed et al., (2017)). We focus on the aspect of marketing ambidexterity that includes exploitation, hoping to find the effect of this considered direction on a company's market performance.

This study tried to address crucial gaps of existing literature, in understanding the impact of BDA-EDCs on market performance, within the specific context of Pakistan, where we interact with firms through a questionnaire. It studied the mediating effect of marketing Ambidexterity in generating these dynamics, with a focus on exploitation, and the impact of Competitor pressure in influencing these dynamics. **_Figure 1_** shows proposed research Framework. The study intends to expand on prior studies by applying the research methodology to smaller organizations and investigating the effect of BDA on profitability in various sized companies and competitive settings. In previous study Adiguzel et al., (2023) future gap is, it would be valuable to apply the same research framework to smaller companies to see if the consequences are similar or different from those in larger firms. In previous study Raguseo et al., (2020) finds that larger firms generally tend to benefit more from BDA investments, it shows that the impact of BDA on profitability varies based on a company's size and the competitive environment in its industry. Addressing these gaps, the present study seeks to offer a complete understanding of BDA-EDCs on market performance, bridging existing knowledge disparities in this domain.

- How do BDA-EDCs and Marketing Ambidexterity Influence Market Performance in firms?
- Does Marketing Ambidexterity (including exploitation) mediate the relationship between BDA-EDCs and Market Performance?
- How does Competitor Pressure moderate the relationship between BDA-EDCs, Marketing Ambidexterity, and Market Performance?

In this study, our primary focus was to conduct an empirical analysis of the impact of BDA-EDCs on the development of marketing ambidexterity, with a particular emphasis on the exploitation element and its subsequent contribution to market overall performance. Our preference of variables stems from a deliberate choice process primarily based on their critical role in influencing a firm's strategic direction and responses to external changes. The study delves into the DCs framework, arguing that BDA-EDCs play a critical role in aligning internal strategies with external needs.

Additionally, this study examines the relationship among BDA and a company's performance in the market, bringing attention to internal operations like marketing ambidexterity (exploitation), a strategic aspect that has been largely ignored in earlier research (2023). We highlight how marketing ambidexterity directly impacts market performance and exhibit the way it connects BDA-EDC to overall market success. The study also discusses the moderating impact of competition pressure, highlighting the importance of contextual factors in assessing a company's efficacy. It means that businesses are more likely to gain advantage from BDA-EDCs in terms of marketing ambidexterity and, subsequently, market performance, mainly when faced with fierce competition.

Moreover, this study targeted on the textile sector in Faisalabad, called the "Manchester of Pakistan," also Chiniot and Lahore Pakistan. It offers a specific context for our investigation. While acknowledging that other industries may be impacted by big data, this targeted approach allows us to gain a nuanced understanding of Big Data and marketing exploitation in the unique dynamics of the textile industry. This deliberate selection aims to contribute depth and specificity to the broader discourse on these variables, shedding light on their implications in a sector renowned for its rapid development within the textile industry.

This research study plan is organized as follows: the first part explains an introduction, then the next part describes Literature, Hypotheses and Framework. Following the literature, the third one explains measures, analysis results, conclusion and a questionnaire that was utilized to gather data.

## LITERATURE REVIEW AND HYPOTHESIS DEVELOPMENT

### Market Performance

A company's good and elevated market performance is described through outperforming competition in marketplace proportion, market share, growth, sales, and product improvement. By means of assisting within the discovery of possible market opportunities, BDAC greatly adds to improving organizational market performance Oluwaseun et al., (2022). In keeping with Yasmin et al., (2020), Big data has a good link with market operations results consisting of market share and sales growth. The influence of BDA-EDC on market overall performance displays its importance. A firm's growth of a strong BDAC has the potential to significantly improve its market performance Olabode et al., (2022); Wamba et al., (2017). This declare remains founded on the notion that BDAC permits companies towards view the market from a unique perspective Olabode et al., (2022).

### Big Data Analytics

We are living in the "Age of data," with new data being generated a supreme and evergrowing rate from all organizations and government agencies Mikalef et al., (2019a). As a result, there has been a lot of buzz, which has led to significant investments by enterprises in their search to find out how they could leverage their data to produce value Mikalef et al., (2019a). The expansion of large datasets rising from numerous resources along with social media, the internet of things, and multimedia applications has increased the significance of big data as a research subject Mohamed et al., (2020). BDA is a broad term for a method that manages, procedures, and evaluates the five Vs of Big data. Big Data consists of an extensive series of information from numerous sources and industries, in each structured and unstructured formats Mikalef et al., (2019a). Through BDAC, big data may be analyzed and customized in line with the specific requirements of a given company Horng et al., (2022). BDA is the application of advanced technologies and analytical techniques to derive significant insights and value from huge amounts of data Mikalef et al., (2019b).

**BDA-Enabled Dynamic Capabilities**

The literature provides several interpretations of Dynamic Capabilities and Big Data. For example, firms with dynamic capabilities can modify, improve, and respond to external developments, offering them a competitive edge and enhanced performance Fosso-Wamba et al., (2019). Fainshmidt et al., (2016) investigate dynamic capacities, with a focus on resource-based and developmental reasons, emphasizing companies' potential to create, extend, or change their asset base on reason. Dynamic capabilities significantly enhance routine business operations in both stable and fast-changing environment Wilhelm et al., (2015). These capabilities not only enhance effectiveness but also make operations more efficient, making understanding crucial for businesses to adapt and perform optimally in changing conditions.

Dynamic capabilities refer to an organization's capacity to adapt and improve, and are characterized by three fundamental components: sensing, seizing, and transforming or reconfiguring Teece, (2014). Sensing involves recognizing external changes and opportunities, Seizing involves effectively capturing these opportunities, and Transforming or Reconfiguring involves altering internal structures to align with new opportunities. These elements enable organizations to adapt and thrive in a changing business landscape. In another study, dynamic capabilities, a key aspect of big data analytics (BDA), involves these three key components. Sensing involves identifying technical options that align with customer prospects, bridging business gaps while seizing allows for efficient resource allocation, capturing emerging business prospects, and maximizing overall value and reconfiguring involves adaptively reorganizing vital capabilities, enabling innovation and market response Saeed et al., (2023), these capabilities, harnessed through BDA, enhance organizational performance by identifying opportunities, strategically allocating resources, and adapting to evolving market dynamics. Big Data Analytics (BDA) is a powerful tool for organizations to process and interpret massive amounts of data, assisting in informed decision-making and identifying trend Fosso-Wamba et al., (2019), this research explores how the application of BDA can improve an organization's dynamic capability. It entails analyzing different types of data, such as five Vs of data Saeed et al., (2023). Mikalef et al., (2019b) investigate dynamic capabilities, emphasizing the importance of a company's ability to sense, coordinate, learn, coordinate, and reconfigure schedules. Firms' ability defined as a BDAC to acquire data to develop insights by successfully organizing and innovate data Mikalef et al., (2019b).

So, the current study views BDA-EDCs as a multiway assembly with three factors generated from big data analytical structures: Sensing, Seizing, And Reconfiguring.

## BDA-Enabled DCs and Marketing Ambidexterity

MA implies to a company's capacity to balance exploration and exploitation strategies in marketing Saeed et al., (2023). Exploration entails searching for and entering new market opportunities, frequently by developing new skills, processes, and strategies while exploitation is the process of optimizing and leveraging existing technologies, processes, and resources in order to achieve valuable results and strengthen the company's position in current markets He et al., (2021). According to another study, exploration involves seeking out new markets, technologies, and capabilities, whereas exploitation focuses on optimizing existing resources and processes for existing markets Saeed et al., (2023). Strategic marketing flexibility is an ability of firms to swiftly adjust their marketing strategies to evolving market conditions, thereby enabling it to explore new opportunities while utilizing existing ones Kouropalatis et al., (2012). To balance exploration and exploitation in marketing efforts, marketing ambidexterity necessitates effective resource allocation and planning. Firms must integrate these processes to successfully pursue both exploration and exploitation, ensuring efficient planning and effective marketing strategy implementation. Marketing ambidexterity implies that a successful company is adept at both exploring new avenues and effectively utilizing what they already must succeed in the market. It is a strategic balance between marketing innovation and optimization He et al., (2021). Usually, firms create value by focusing on one compliant processes (exploration or exploitation) Vorhies et al., (2011). According to the study Yamakawa et al., (2011), exploitation associations may provide more obvious and instant benefits to the main firm than exploration alliances. In this study, particularly focusing on exploitation.

In today's business landscape, BDA is regarded as a critical tool for improvement, competition, and productivity Saeed et al., (2023). BDA-EDCs have a direct relationship with marketing ambidexterity. Study Lies, (2019) highlights the integration of big data and marketing insights, driving to a move from outdated outbound marketing to more personalized inbound marketing.

### BDA-Enabled DCs and Market Performance

As theorized within the earlier segments, BDA-EDCs are specifically associated with marketing ambidexterity, and MA is associated with market performance Shafique et al., (2023). Taking after argument, Fosso-Wamba et al., (2019), an indirect association is expected with this research between BDA-EDC and firm's market performance; hence, marketing ambidexterity act as a mediator between the BDA-EDC and market performance. Particularly, firms with high marketing ambidexterity and BDA, they are more viable, and these capabilities have a positive effect on market performance. Dynamic capabilities refer to an enterprise's capacity to efficiently construct and renew resources and assets, adapting to market changes, and can be categorized into three clusters: sensing, seizing, and transforming, which include continuous reestablishment and reconfiguration Teece, (2014). The literature suggests that firms must enhance their big data analytics capabilities to effectively utilize this technology and achieve performance improvements Mikalef et al., (2019a).

On the potential building of BDA-enabled dynamic capabilities to reshape firms' operations and performance, we expect a positive relationship, where firms effectively utilizing BDA for adaptation, innovation, and strategic decision-making will demonstrate better market performance.

$H_1$: BDA-EDCs positively influence Firm Performance, particularly in terms of Market Performance.

### Marketing Ambidexterity and Market Performance

When a company invests more resources in big data development, marketing competencies are boosted while organizations apply big data to marketing, which leads to improved corporate overall performance Horng et al., (2022). BDA-enabled DCs boost a company's market and financial performance by improving marketing ambidexterity Saeed et al., (2023). The study helps to understand how strategically ambidextrous firms can achieve superior performance outcomes in dynamic market conditions by effectively balancing flexibility and commitment Kouropalatis et al., (2012). According to the study (Yamakawa et al., 2011), firms that form more exploitation alliances (rather than exploration alliances) perform better in the short term. This study (Mizik & Jacobson, 2003) suggests that exploitation and exploration are both important to improve firm market performance. This implies that new capabilities may be required to transform novel ideas into advanced outcomes, which can ultimately benefit the organization's performance (Ferreira et al., 2020). The study Mehrabi et al., (2019) investigates how the level and balance of ambidexterity in customer management and new product development affects a firm's overall performance. Balanced ambidexterity refers to an equal pursuit of exploration and exploitation, whereas combined ambidexterity refers to increased levels of both exploration and exploitation.

## Mediating Role of Marketing Ambidexterity

The integration of BDA-EDCs enables firms to leverage available data resources systematically, refine existing marketing strategies, and optimize operations. Exploration alliances are more advantageous in high-growth businesses, whereas exploitation alliances are more profitable in low-growth industries for generating short-term financial rewards Yamakawa et al., (2011). This study highlights the significance of growth in industry in determining the impact of exploitation. Younger enterprises gain more from exploitative alliances, as they sometimes lack internal resources and expertise Yamakawa et al., (2011). Exploitation alliances allow them to make better use of existing assets.

BDA-enabled DCs enable firms to effectively adapt to external changes, enhancing marketing ambidexterity (exploitation) and achieving a sustained competitive advantage. Firms with strong Dynamic Capabilities expected to improve the scale of marketing ambidexterity, thus positively contributing to a firm's marketing ambidexterity (2023). So,

$H_2$: Marketing Ambidexterity (exploitation) mediates the relationship between BDAEDC and Market Performance.

**Competitor Pressure as Moderator**

The term "competitive pressure" refers to the observed pressure from competitors that forces a firm to embrace modern technology to gain a competitive edge ( 2021).

Previous empirical research showed the significant impact of competitive pressure on BDA adoption intentions Agrawal, (2015); Alaskar et al., (2021). Furthermore, no marketing strategy is developed without thinking about competitors Chabowski et al., (2011). BDA entails organizing, processing, and studying massive amounts of data so that you can gain valuable insights for decision-making and competitive advantage Saeed et al., (2023). Competitive advantage is crucial for achieving advanced company performance, value creation, and long-term maintenance Teece, (2014). When competition is extreme, companies need to renew in each product and processes, explore new markets, and consider how they'll distinguish themselves from competitors Zahra, (1993). BDA allows organizations to make real-time data-driven decisions, enhance operational efficiency, create new products, reduce costs, and gain a competitive advantage Mikalef et al., (2019b). The research results highlight the significance of competitive pressure in influencing the adoption of technology, especially social media marketing, by SMEs, it mentions that competitive pressure arises from the risk of losing a competitive edge in the industry, which prompts firms to embrace technological changes Ali Abbasi et al., (2022).

Firms that prioritize differentiation have to pursue more exploration alliances, whereas firms that prioritize cost leadership should pursue more exploitation alliances Yamakawa et al., (2011). According to one study Ali Abbasi et al., (2022), adopting technology such as social media marketing, can adjust the competitive dynamics within an industry; firms that embody these changes can gain an advantage, subsequently influencing industry competition.

As highlighted, in the bustling industrial landscape of Faisalabad, Chiniot, Lahore Pakistan, this study proposed that competitor pressure acts as a crucial moderator. specifically, we assume that firms will experience a stronger positive effect on market performance from BDA-enabled dynamic capabilities when they face higher levels of competition, show the strategic adaptability of dynamic capabilities in response to competitive pressures. So, the current research hunt for to support the findings of prior research by empirically evaluating competitive pressure as a moderator in the relationship between BDA-EDC and market performance; thus, the following third hypotheses are proposed:

$H_3$: Competitor Pressure moderates the relationship between BDA-EDC and Market Performance.

These hypotheses form a comprehensive framework for study's research, encompassing the interplay between BDA-EDC, marketing ambidexterity (Exploitation), competitor pressure, and their ultimate influence on Market performance. They provide a structured path to investigate the complex relationships within our research context, enabling us to contribute valuable insights to the field.

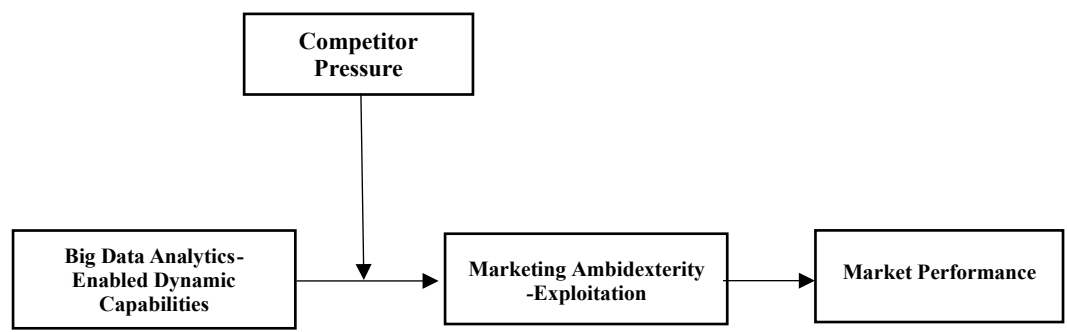

*Figure 1. Research Framework*

## RESEARCH METHODOLOGY

To evaluate the hypothesis, the current study has used one independent sample. The survey method become employed to achieve data. The research purpose was to evaluate the effect of big data on marketing ambidexterity (particularly exploitation) then market performance, so CEOs and executives had been contacted and who understand firm operations, for data collection via personal and professional connections.

The study is to target a representative sample from the retail firms of Faisalabad, Chiniot and Lahore Pakistan of textile sector. To capture multiple viewpoints, the selection criteria included aspects such as company size, years in operation, and technological usage. We have collected data from 303 retail firms in the textile industry who deal directly with customers. This procedure entailed contacting recognized textile companies. For data, one of the researchers approached numerous textile firms in Faisalabad, Chiniot and Lahore Pakistan and request them to add their contribution in the study. The survey is distributed along with a cover letter. The cover letter contains all the information about the topic under investigation, including details about the survey's conductors and the purpose of research.

Respondents included major decision makers such as managers and executives who are knowledgeable about the company's operations. Depending on the preferences of the participating companies, the survey is distributed to them through online, email and face-to-face interactions. Firms selected for the survey are those directly dealing with customers in the textile industry located in Chiniot (33), Faisalabad (187), and Lahore (83).

## Data Analysis Tools and Techniques

For hypotheses testing, the research used PLS-SEM. PLS-SEM was chosen for its capacity to tolerate measurement mistakes during unknown indicator aggregation, resulting in a more accurate representation of latent variables. This analysis will yield correlation coefficients across variables and help to understand the linkages between BDA-enabled Dynamic Capabilities, Marketing Ambidexterity, Competitor Pressure, and Market Performance. A detailed questionnaire is created to collect data from certain companies. The questionnaire was delivered in its English language. The sample collected from textile firms. BDA-EDCs, Marketing Ambidexterity (Exploitation), Competitor Pressure, and Market Performance are all covered in the questionnaire. To examine the impact of BDA on a firm's ability to adapt to change and optimize data resources, questions were addressed. The questionnaire also investigates the effect of Marketing Ambidexterity (particularly exploitation), which mediates the relationship between variables.

## Measures

Pre-developed questionnaires are used to measure components on a scale. BDA-EDC; Sens, Seiz and Reco, calculated on a 10-item scale altered from (2019; 2023). Marketing Ambidexterity particularly exploitation (Explt), used 4 item scale of (2023; 2011) to measure it. Market performance (Mktp) evaluated by a 4-item scale altered from (2023; 2012). Respondents asked to rate the improvement in market performance compared to competitors. The Competitor Pressure (COMPRE) scale applied on this questionnaire has been adapted from a previously established source (2013; 2020), the scale has been thoughtfully tailored to the specific context of this research project. Respondents asked to consider your firm's role and the prevailing beliefs within your industry while responding to the statements.

Firms indicated the extent to which they use BDA-EDC, use analytics to modify existing marketing procedures in comparison to their competitors, use analytics to achieve the following market performance statements over the last three years, and competitor pressure. They scored each statement from Strongly Disagree to 5 Strongly Agree depending on how well it matched their firm's experiences. For the following questionnaire statements see *Table 6 (Appendix).*

## ANALYSIS AND RESULTS

The PLS model was used to evaluate the hypothesized framework. The most common motives for employing PLS-SEM are small sample size, formative measures, and prediction Hair et al., (2012). A PLS is usually analyzed in two levels. The measurement model is the first model, tested by appearing validity and reliability analyses of each of the statements inside the model Hulland, (1999). The structure model is second model, tested by estimating the paths between the items in the model, determining their importance and the model's predictive ability; the reason for using PLS-SEM was because it is a beneficial tool for marketing studies Hulland, (1999). PLS was employed for modeling the path and analyzing data.

In addition, PLS-SEM reduces complexity and uncertainty via the use of flexible assumptions to estimate models, resulting in increased theoretical consistency (Hair et al.,(2011). A preliminary analysis was undertaken earlier than the main analysis to prevent issues with multicollinearity, missing values, and outliers.

### Measurement Model

Reliability and validity are essential elements in evaluating studies quality, making sure consistency and accuracy in measurement. Reliability, gauged by using measures like Cronbach's alpha and composite reliability, preferably needs to exceed 0.70 to be considered acceptable Robert A. Peterson, (1994). Above 0.9 Cronbach's alpha values show excellent reliability, while those above 0.8 and 0.7 signify good and acceptable ranges, respectively Wadkar et al., (2016). Higher Cronbach's alpha scores mean stronger internal consistency among scale items. In the present study, all constructs displayed CR values of more than 0.6, signifying satisfactory consistency and reliability among scale items.

Convergent validity, assessed through the average Variance Extracted (AVE), validates the relationship among constructs. 0.50 average variance extracted is accepted Hair et al., (2011). In this study, all values surpassed this threshold, affirming convergent validity. Discriminant validity, alternatively, examines the uniqueness of constructs. This was evaluated through factors which include factor loading, HTMT ratio, and the Fornell & Larcker criterion. *Table 2* shows the HTMT ratio. *Table 1* shows the CR and validity, illustrating the Cronbach's alpha, CR, CR (rho_c), and Average Variance Extracted for each construct. These values collectively verify the reliability and validity of the instruments applied within the study.

*Table 1: Construct Reliability and Validity*

|  | Cronbach's alpha | Composite Reliability (rho_a) | Composite Reliability (rho_c) | Average Variance Extracted (AVE) |
|---|---|---|---|---|
| **BDA-EDC** | 0.907 | 0.908 | 0.924 | 0.553 |
| **CP** | 0.877 | 0.933 | 0.910 | 0.640 |
| **MA** | 0.913 | 0.919 | 0.938 | 0.792 |
| **MP** | 0.846 | 0.879 | 0.894 | 0.678 |

The Heterotrait-Monotrait Ratio (HTMT) matrix serves as an essential tool in assessing discriminant validity, with values preferably below 1.00 Henseler et al., (2015). This ratio suggests the degree of similarity among latent variables, with values smaller than one signifying established discriminant validity. *Table 2* illustrates HTMT ratios for each assemble pair, demonstrating that all ratios fall within acceptable ranges, affirming the presence of DV across all constructs.

*Table 2: Heterotrait-Monotrait Ratio (HTMT)*

|  | *Heterotrait-Monotrait Ratio (HTMT)* |
|---|---|
| **CP <-> BDA-EDC** | 0.413 |
| **MA <-> BDA-EDC** | 0.383 |
| **MA <-> CP** | 0.446 |
| **MP <-> BDA-EDC** | 0.647 |
| **MP <-> CP** | 0.527 |
| **MP <-> MA** | 0.365 |

## Hypothesis Testing

Studies has established that Structural Equation Modeling (SEM) is an effective approach for confirming theoretically relationship Akter et al., (2017), however, acquiring similar results using numerous methods could increase confidence in the stability of these findings Hultman et al., (2021). Results show a positive relation between BDA-EDC and marketing ambidexterity, furthermore, BDA-EDC and marketing ambidexterity have positive links with firm performance.

The first hypothesis was developed that, "BDA-EDC positively affect firm performance, especially in terms of market performance." So, this hypothesis of the study examined through statistical software, and regression analysis was used to check the impact of BDA-EDC on company's market performance. The test results indicated that (b=0.242, t=3.843, p= 0.000) this means that that Beta is saying that one unit change in BDA-EDC will create 0.242-unit change in market performance. A p-value less than

0.05 is c considered significant, t-value greater than +2 are considered acceptable & significant. P-value (0.000) for $H_1$ is less than 0.05, indicating statistical significance. t value (3.834) is greater than +2, confirming the significance of this relationship. So, those results show the connection between BDA and firm performance is significant. big data related results that support previous studies include the ones by Dubey et al., (2022); Fosso-Wamba et al., (2019); Mikalef et al., (2019b); Saeed et al., (2023). The second hypothesis is "marketing Ambidexterity (exploitation) mediates the relationship between BDA-EDC and market performance." The test results indicated that (b= 0.343, t= 5.483, p= 0.000) this means that that Beta is saying that one unit change in marketing ambidexterity (exploitation) will create 0.343-unit change in BDA-EDC and market performance. The p-value (0.000) and t statistics value (5.483) for $H_2$ each imply statistical significance. T values greater than +2 are considere acceptable & significant. The P-value (0.000) for the second hypothesis is less than 0.05, indicating significance. t statistics value (5.483) is more than +2, confirming the significance of this relationship. So, those results show that $H_2$ is accepted, suggesting that marketing Ambidexterity, specifically exploitation, mediates the relationship.

*Table 3: Path coefficients*

|  | Beta coefficient | Sample Mean | Standard deviation | T statistics | P values |
|---|---|---|---|---|---|
| $H_1$ | 0.242 | 0.244 | 0.063 | 3.843 | 0.000 |
| $H_2$ | 0.343 | 0.348 | 0.063 | 5.483 | 0.000 |
| $H_3$ | -0.129 | -0.126 | 0.055 | 2.356 | 0.019 |

And the third hypothesis is that "Competitor pressure moderates the relation between BDA and market performance." The test results indicated that (b= -0.129, t= 2.356, p= 0.019) which means that Beta is saying that one unit change in competitor pressure will create -0.129-unit change in BDA-EDC. The p-value (0.019) and t statistics value (2.356) of $H_3$ show statistical significance. A negative beta coefficient suggests that as competitor pressure increases, market performance tends to lower. This could mean that higher levels of competition in the market may have an adverse impact on the overall performance, leading to lower market performance outcomes. So, $H_3$ is accepted, indicating that Competitor pressure moderates the relationship among BDA-EDCs and market performance.

### *Table 4: Outer Model*

| | Beta | Sample Mean | Standard deviation | t statistics | P values |
|---|---|---|---|---|---|
| **BDA-EDC1 <- BDA-EDC** | 0.724 | 0.718 | 0.051 | 14.128 | 0.000 |
| **BDA-EDC2 <- BDA-EDC** | 0.705 | 0.701 | 0.047 | 14.944 | 0.000 |
| **BDA-EDC3 <- BDA-EDC** | 0.500 | 0.501 | 0.067 | 7.439 | 0.000 |
| **BDA-EDC4 <- BDA-EDC** | 0.701 | 0.698 | 0.051 | 13.801 | 0.000 |
| **BDA-EDC5 <- BDA-EDC** | 0.834 | 0.831 | 0.029 | 28.596 | 0.000 |
| **BDA-EDC6 <- BDA-EDC** | 0.846 | 0.843 | 0.029 | 29.284 | 0.000 |
| **BDA-EDC7 <- BDA-EDC** | 0.843 | 0.840 | 0.027 | 31.244 | 0.000 |
| **BDA-EDC8 <- BDA-EDC** | 0.684 | 0.684 | 0.042 | 16.301 | 0.000 |
| **BDA-EDC9 <- BDA-EDC** | 0.735 | 0.734 | 0.040 | 18.478 | 0.000 |
| **BDA-EDC10 <- BDA EDC** | 0.795 | 0.792 | 0.037 | 21.226 | 0.000 |
| **CP1 <- CP** | 0.917 | 0.917 | 0.018 | 51.871 | 0.000 |
| **CP2 <- CP** | 0.888 | 0.885 | 0.025 | 35.566 | 0.000 |
| **CP3 <- CP** | 0.873 | 0.873 | 0.035 | 24.848 | 0.000 |
| **CP4 <- CP** | 0.706 | 0.701 | 0.066 | 10.625 | 0.000 |
| **CP5 <- CP** | 0.880 | 0.879 | 0.025 | 34.710 | 0.000 |
| **CP6 <- CP** | 0.418 | 0.413 | 0.081 | 5.191 | 0.000 |
| **MA1 <- MA** | 0.897 | 0.896 | 0.021 | 42.078 | 0.000 |
| **MA2 <- MA** | 0.865 | 0.864 | 0.029 | 29.770 | 0.000 |
| **MA3 <- MA** | 0.904 | 0.904 | 0.018 | 49.582 | 0.000 |
| **MA4 <- MA** | 0.893 | 0.892 | 0.022 | 40.379 | 0.000 |
| **MP1 <- MP** | 0.730 | 0.720 | 0.066 | 10.987 | 0.000 |
| **MP2 <- MP** | 0.806 | 0.809 | 0.038 | 21.048 | 0.000 |
| **MP3 <- MP** | 0865 | 0.860 | 0.033 | 25.861 | 0.000 |
| **MP4 <- MP** | 0.885 | 0.881 | 0.026 | 33.710 | 0.000 |
| **CP x BDA-EDC -> CP x BDA-EDC** | 1.000 | 1.000 | 0.000 | n/a | n/a |

## THEORETICAL IMPLICATION

Research contributes to the literature of Big Data and marketing ambidexterity in DCs theory. We comprehensively examined the implications of our results within the setting of the research goals and theoretical framework. Particularly, this research shows the impact of BDA-EDC in upgrading firm performance, especially in terms of market performance. Our examination into the mediating impact of marketing ambidexterity shed light on the components through which BDA-EDCs influence market performance. The results of research show the marketing ambidexterity determined from BDA incorporates a positive effect on performance of companies, back by earlier studies Fosso-Wamba et al., (2019); Yasmin et al., (2020); Saeed et al., (2023).

Furthermore, by addressing the moderating role of competitor pressure, we gained useful insights into the contextual factors that affect the relation between big data marketing ambidexterity, and market performance. The conclusions underline the significance of studying aspects in understanding the dynamics of organizational capabilities and their impact on performance outcomes.

## LIMITATIONS AND FUTURE DIRECTIONS

Limitations of research are Firstly; Future work should include both objective and subjective performance assessments to identify any differential impacts. Secondly, our research focused exclusively on the textile sector in Pakistan, which may limit the universality of the results to other industries or geographical perspectives. In this analysis, competitor pressure moderated the BDA-enabled DCs nexus only in a few Pakistani cities. Future studies should investigate the different effects of competitor pressure in other cities in Pakistan and other countries. Furthermore, researching this phenomenon using qualitative approaches, such as case studies, would be interesting.

Additionally, the cross-sectional nature of data collection restricts this study's capability to form fundamental relationships. Further research actions could benefit from studies to provide more understanding of the changing aspects between BDA-EDC, marketing ambidexterity, competitor pressure, and firm's market performance across different industries and over time. *Graphs 1* and *2* show the graphical analysis of PLS-SEM, factor analysis and path analysis simultaneously.

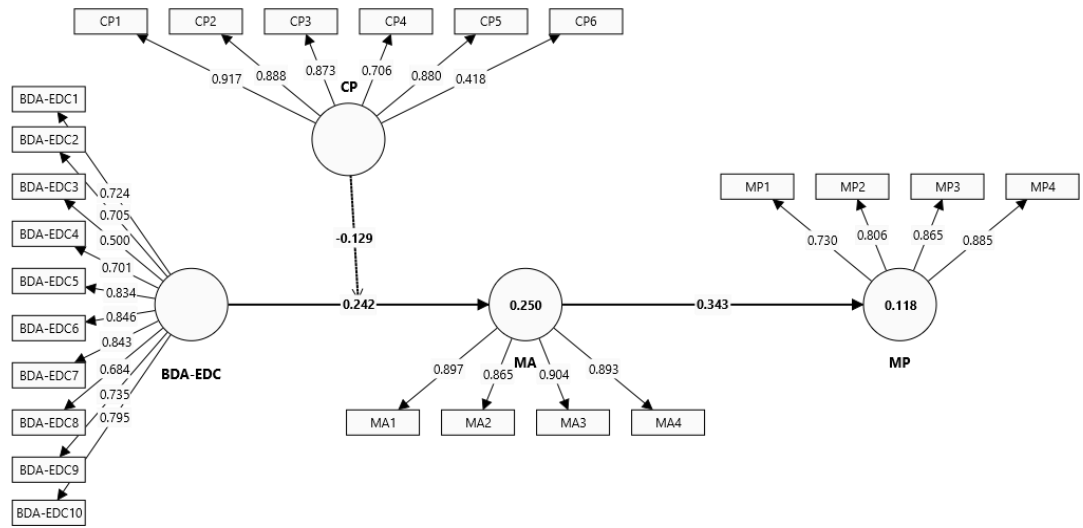

*Graph 1: Factor Analysis*

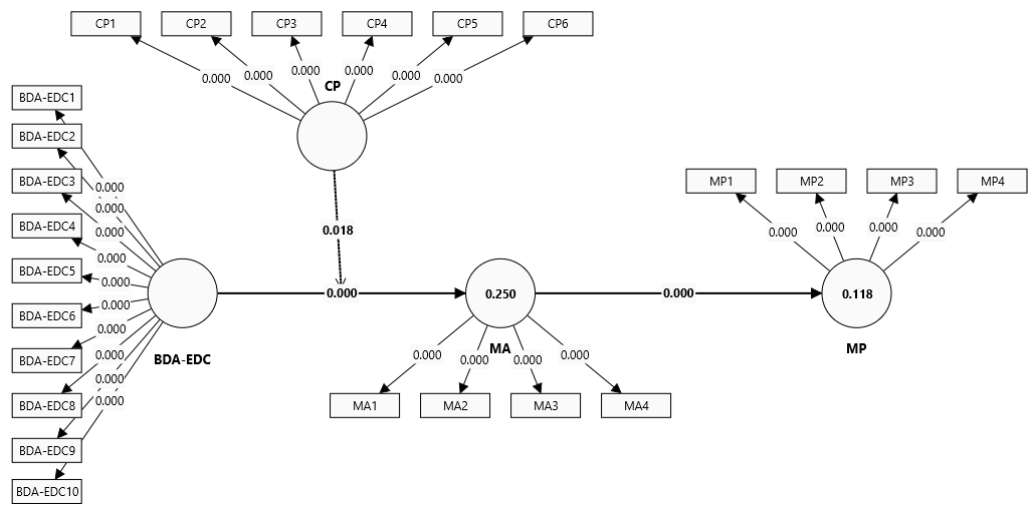

*Graph 2: Path Analysis*

## MANAGERIAL IMPLICATION

The research carries notable implications for managerial practice within the textile industry. Firstly, it underscores the importance of strategic investments in Big Data Analytics (BDA), particularly aimed at fostering BDA-enabled Dynamic Capabilities (DCs). It is imperative for firms to recognize that the true value of BDA-enabled DCs lies in their ability to ensure perfect and uninterrupted flow of information over time, enabling organizations to adapt swiftly to evolving market dynamics. The results of the study have huge practical impact for various industries developing Big Data Analytics. By improving Big Data Analytics, managers could better serve consumer needs, increase sales with quality information Akter et al., (2016).

Furthermore, noticing feature is that this study is conducted within the context of Pakistan, where literature on such topics within developing countries is limited. Therefore, the implications extend beyond individual firms to the broader textile industry, offering valuable guidance for organizations seeking to overcome challenges within their firms and capitalize on emerging opportunities.

## CONCLUSION

The research was done to research how BDA impacts a firm's market performance in the textile industry. Research proves the positive relation between big data and performance. Previous research has confirmed that big data strategies have a significant impact on firm's market performance Horng et al., (2022). Second, this study hypothesis suggested that BDA-EDC empowers firms to effectively harness market knowledge, leading to enhanced marketing ambidexterity and consequently, improved market performance. Through analysis of diverse samples, our findings affirm a positive association between BDA-EDC and marketing ambidexterity, particularly in the dimension of marketing exploitation. Furthermore, the study reveals that the adept utilization of marketing exploitation correlates positively with firms' market performance.

To examine the connection between BDA-EDC and market performance, this study delved into the moderating effect of competitor pressure on these relationships. The results suggest that among varying levels of competitor pressure, firms can leverage BDAEDC to modify their marketing strategies, leveraging insights concerning markets and products to attain a competitive edge and elevate overall performance. This investigation contributes to the discussion on Big Data and marketing ambidexterity within the framework of Dynamic Capabilities theory, explaining the numerous ways in which marketing ambidexterity, derived from BDA-EDC, positively impacts market performance in the textile industry.

## REFERENCES

Agrawal, K. P. (2015). Investigating the determinants of Big Data Analytics (BDA) adoption in asian emerging economies. *2015 Americas Conference on Information Systems, AMCIS 2015*, 1–18. https://doi.org/10.5465/ambpp.2015.11290abstract

Ahmed, M., Haider, G., & Zaman, A. (2017). Detecting structural change with heteroskedasticity. *Communications in Statistics - Theory and Methods*, *46*(21), 10446–10455. https://doi.org/10.1080/03610926.2016.1235200

Aj, D. B., Ashraf, S., Muhammad, W., & Gulfam, K. (2022). *Impact of Remittances on Socio-Economic Condition of Rural Families of*. *7*(1), 1–13.

Akter, S., Fosso Wamba, S., & Dewan, S. (2017). Why PLS-SEM is suitable for complex modelling? An empirical illustration in big data analytics quality. *Production Planning and Control*, *28*(11–12), 1011–1021. https://doi.org/10.1080/09537287.2016.1267411

Akter, S., & Wamba, S. F. (2016). Big data analytics in E-commerce: a systematic review and agenda for future research. *Electronic Markets*, *26*(2), 173–194. https://doi.org/10.1007/s12525-0160219-0

Akter, S., Wamba, S. F., Gunasekaran, A., Dubey, R., & Childe, S. J. (2016). How to improve firm performance using big data analytics capability and business strategy alignment? *International Journal of Production Economics*, *182*, 113–131. https://doi.org/10.1016/j.ijpe.2016.08.018

Alaskar, T. H., Mezghani, K., & Alsadi, A. K. (2021). Examining the adoption of Big data analytics in supply chain management under competitive pressure: evidence from Saudi Arabia. *Journal of Decision Systems*, *30*(2–3), 300–320. https://doi.org/10.1080/12460125.2020.1859714

Ali Abbasi, G., Abdul Rahim, N. F., Wu, H., Iranmanesh, M., & Keong, B. N. C. (2022). Determinants of SME's Social Media Marketing Adoption: Competitive Industry as a Moderator. *SAGE Open*, *12*(1). https://doi.org/10.1177/21582440211067220

Bahrami, M., & Shokouhyar, S. (2022). The role of big data analytics capabilities in bolstering supply chain resilience and firm performance: a dynamic capability view. *Information Technology and People*, *35*(5), 1621–1651. https://doi.org/10.1108/ITP-01-2021-0048

Chabowski, B. R., Mena, J. A., & Gonzalez-Padron, T. L. (2011). The structure of sustainability research in marketing, 1958-2008: A basis for future research opportunities. *Journal of the Academy of Marketing Science*, *39*(1), 55–70. https://doi.org/10.1007/s11747-010-0212-7

Côrte-Real, N., Ruivo, P., Oliveira, T., & Popovič, A. (2019). Unlocking the drivers of big data analytics value in firms. *Journal of Business Research*, *97*(January), 160–173. https://doi.org/10.1016/j.jbusres.2018.12.072

Dubey, R., Bryde, D. J., Dwivedi, Y. K., Graham, G., & Foropon, C. (2022). Impact of artificial intelligence-driven big data analytics culture on agility and resilience in humanitarian supply chain: A practice-based view. *International Journal of Production Economics*, *250*(March), 108618. https://doi.org/10.1016/j.ijpe.2022.108618

Fainshmidt, S., Pezeshkan, A., Lance Frazier, M., Nair, A., & Markowski, E. (2016). Dynamic Capabilities and Organizational Performance: A Meta-Analytic Evaluation and Extension. *Journal of Management Studies*, *53*(8), 1348–1380. https://doi.org/10.1111/joms.12213

Ferreira, J., Coelho, A., & Moutinho, L. (2020). Dynamic capabilities, creativity and innovation capability and their impact on competitive advantage and firm performance: The moderating role of entrepreneurial orientation. *Technovation*, *92–93*(July 2018), 102061. https://doi.org/10.1016/j.technovation.2018.11.004

Fosso-Wamba, S., Dubey, R., Gunasekaran, A., & Akter, S. (2019). *"The Performance Effects of Big Data Analytics and*.

Fosso Wamba, S., Akter, S., Edwards, A., Chopin, G., & Gnanzou, D. (2015). How "big data" can make big impact: Findings from a systematic review and a longitudinal case study. *International Journal of Production Economics*, *165*(January), 234–246. https://doi.org/10.1016/j.ijpe.2014.12.031

Gaur, L., Afaq, A., Solanki, A., Singh, G., Sharma, S., Jhanjhi, N. Z., My, H. T., & Le, D. N. (2021). Capitalizing on big data and revolutionary 5G technology: Extracting and visualizing ratings and reviews of global chain hotels. *Computers and Electrical Engineering*, *95*(September 2020), 107374. https://doi.org/10.1016/j.compeleceng.2021.107374

Hair, J. F., Ringle, C. M., & Sarstedt, M. (2011). PLS-SEM: Indeed a silver bullet. *Journal of Marketing Theory and Practice*, *19*(2), 139–152. https://doi.org/10.2753/MTP1069-6679190202

Hair, J. F., Sarstedt, M., Pieper, T. M., & Ringle, C. M. (2012). The Use of Partial Least Squares Structural Equation Modeling in Strategic Management Research: A Review of Past Practices and Recommendations for Future Applications. *Long Range Planning*, *45*(5–6), 320–340. https://doi.org/10.1016/j.lrp.2012.09.008

He, P., Pei, Y., Lin, C., & Ye, D. (2021). Ambidextrous marketing capabilities, exploratory and exploitative market-based innovation, and innovation performance: an empirical study on china's manufacturing sector. *Sustainability (Switzerland)*, *13*(3), 1–21. https://doi.org/10.3390/su13031146

Henseler, J., Ringle, C. M., & Sarstedt, M. (2015). A new criterion for assessing discriminant validity in variance-based structural equation modeling. *Journal of the Academy of Marketing Science*, *43*(1), 115–135. https://doi.org/10.1007/s11747-014-0403-8

Horng, J. S., Liu, C. H., Chou, S. F., Yu, T. Y., & Hu, D. C. (2022). Role of big data capabilities in enhancing competitive advantage and performance in the hospitality sector: Knowledge-based dynamic capabilities view. *Journal of Hospitality and Tourism Management*, *51*(May 2021), 22–38. https://doi.org/10.1016/j.jhtm.2022.02.026

Hsu, C. C., Tan, K. C., Zailani, S. H. M., & Jayaraman, V. (2013). Supply chain drivers that foster the development of green initiatives in an emerging economy. *International Journal of Operations and Production Management*, *33*(6), 656–688. https://doi.org/10.1108/IJOPM-10-2011-0401

Hulland, J. (1999). Use of partial least squares (PLS) in strategic management research: A review of four recent studies. *Strategic Management Journal*, *20*(2), 195–204. https://doi.org/10.1002/(sici)10970266(199902)20:2<195::aid-smj13>3.0.co;2-7

Hultman, M., Iveson, A., & Oghazi, P. (2021). The Information Paradox in Internationalization: Can ignorance ever be bliss? Evidence from emerging market SME managers. *Journal of Business Research*, *131*, 268–277. https://doi.org/10.1016/j.jbusres.2021.03.043

Kouropalatis, Y., Hughes, P., & Morgan, R. E. (2012). Pursuing "flexible commitment" as strategic ambidexterity: An empirical justification in high technology firms. *European Journal of Marketing*, *46*(10), 1389–1417. https://doi.org/10.1108/03090561211248099

Lies, J. (2019). Marketing Intelligence and Big Data: Digital Marketing Techniques on their Way to Becoming Social Engineering Techniques in Marketing. *International Journal of Interactive Multimedia and Artificial Intelligence*, *5*(5), 134. https://doi.org/10.9781/ijimai.2019.05.002

Lutfi, A., Alrawad, M., Alsyouf, A., Almaiah, M. A., Al-Khasawneh, A., Al-Khasawneh, A. L., Alshira'h, A. F., Alshirah, M. H., Saad, M., & Ibrahim, N. (2023). Drivers and impact of big data analytic adoption in the retail industry: A quantitative investigation applying structural equation modeling. *Journal of Retailing and Consumer Services*, *70*(September 2022), 103129. https://doi.org/10.1016/j.jretconser.2022.103129

Manyika, J., Chui Brown, M., B. J., B., Dobbs, R., Roxburgh, C., & Hung Byers, A. (2011). Big data:

The next frontier for innovation, competition and productivity. *McKinsey Global Institute*, *June*, 156. https://bigdatawg.nist.gov/pdf/MGI_big_data_full_report.pdf

Mehrabi, H., Coviello, N., & Ranaweera, C. (2019). Ambidextrous marketing capabilities and performance: How and when entrepreneurial orientation makes a difference. *Industrial Marketing Management*, *77*(January 2018), 129–142. https://doi.org/10.1016/j.indmarman.2018.11.014

Mikalef, P., Boura, M., Lekakos, G., & Krogstie, J. (2019a). Big data analytics and firm performance: Findings from a mixed-method approach. *Journal of Business Research*, *98*(January), 261–276. https://doi.org/10.1016/j.jbusres.2019.01.044

Mikalef, P., Boura, M., Lekakos, G., & Krogstie, J. (2019b). Big Data Analytics Capabilities and Innovation: The Mediating Role of Dynamic Capabilities and Moderating Effect of the Environment. *British Journal of Management*, *30*(2), 272–298. https://doi.org/10.1111/14678551.12343

Mithas, S., Ramasubbu, N., & Sambamurthy, V. (2011). Institutional Knowledge at Singapore Management University. *Research Collection School Of Information Systems*, *3*(1), 237–256.

Mizik, N., & Jacobson, R. (2003). Trading off between value creation and value apropriation. *Journal of Marketing*, *67*(January), 63–76.

Mohamed, A., Najafabadi, M. K., Wah, Y. B., Zaman, E. A. K., & Maskat, R. (2020). The state of the art and taxonomy of big data analytics: view from new big data framework. In *Artificial Intelligence Review* (Vol. 53, Issue 2). Springer Netherlands. https://doi.org/10.1007/s10462-01909685-9

Olabode, O. E., Boso, N., Hultman, M., & Leonidou, C. N. (2022). Big data analytics capability and market performance: The roles of disruptive business models and competitive intensity. *Journal of Business Research*, *139*(November 2021), 1218–1230. https://doi.org/10.1016/j.jbusres.2021.10.042

Raguseo, E., Vitari, C., & Pigni, F. (2020). Profiting from big data analytics: The moderating roles of industry concentration and firm size. *International Journal of Production Economics*, *229*. https://doi.org/10.1016/j.ijpe.2020.107758

Rialti, R., Zollo, L., Ferraris, A., & Alon, I. (2019). Big data analytics capabilities and performance: Evidence from a moderated multi-mediation model. *Technological Forecasting and Social Change*, *149*(June), 119781. https://doi.org/10.1016/j.techfore.2019.119781

Riva, F., & Gani, M. O. (2020). Effect of Customer and Competitor Pressure on Environmental Performance of Upscale Hotels in Bangladesh. *Jahangirnagar University Journal of Business Research*, *21*(June 2020), 193–212. https://www.researchgate.net/publication/349486027

Robert A. Peterson. (1994). A Meta-Analysis of Cronbach's Coefficient Alpha. *Journal of Consumer Research*, *21*(2), 381–391.

Saeed, M., Adiguzel, Z., Shafique, I., Kalyar, M. N., & Abrudan, D. B. (2023). Big data analyticsenabled dynamic capabilities and firm performance: examining the roles of marketing ambidexterity and environmental dynamism. *Business Process Management Journal*, *29*(4), 1204–1226. https://doi.org/10.1108/BPMJ-01-2023-0015

Teece, D. J. (2014). The foundations of enterprise performance: Dynamic and ordinary capabilities in an (economic) theory of firms. *Academy of Management Perspectives*, *28*(4), 328–352. https://doi.org/10.5465/amp.2013.0116

Vorhies, D. W., Orr, L. M., & Bush, V. D. (2011). Improving customer-focused marketing capabilities and firm financial performance via marketing exploration and exploitation. *Journal of the Academy of Marketing Science*, *39*(5), 736–756. https://doi.org/10.1007/s11747-010-0228-z

Wadkar, S. K., Singh, K., Chakravarty, R., & Argade, S. D. (2016). Assessing the Reliability of Attitude Scale by Cronbach's Alpha. *Journal of Global Communication*, *9*(2), 113. https://doi.org/10.5958/0976-2442.2016.00019.7

Wamba, S. F., Gunasekaran, A., Akter, S., Ren, S. J. fan, Dubey, R., & Childe, S. J. (2017). Big data analytics and firm performance: Effects of dynamic capabilities. *Journal of Business Research*, *70*, 356–365. https://doi.org/10.1016/j.jbusres.2016.08.009

Wang, N., Liang, H., Zhong, W., Xue, Y., & Xiao, J. (2012). Resource structuring or capability building? An empirical study of the business value of information technology. *Journal of Management Information Systems*, *29*(2), 325–367. https://doi.org/10.2753/MIS0742-1222290211

Wilhelm, H., Schlömer, M., & Maurer, I. (2015). How dynamic capabilities affect the effectiveness and efficiency of operating routines under high and low levels of environmental dynamism. *British Journal of Management*, *26*(2), 327–345. https://doi.org/10.1111/1467-8551.12085

Yamakawa, Y., Yang, H., & Lin, Z. (2011). Exploration versus exploitation in alliance portfolio: Performance implications of organizational, strategic, and environmental fit. *Research Policy*, *40*(2), 287–296. https://doi.org/10.1016/j.respol.2010.10.006

Yasmin, M., Tatoglu, E., Kilic, H. S., Zaim, S., & Delen, D. (2020). Big data analytics capabilities and firm performance: An integrated MCDM approach. *Journal of Business Research*, *114*(March), 1–15. https://doi.org/10.1016/j.jbusres.2020.03.028

Zahra, S. A. (1993). Environment, corporate entrepreneurship, and financial performance: A taxonomic approach. *Journal of Business Venturing*, *8*(4), 319–340. https://doi.org/10.1016/08839026(93)90003-N

Zhang, D., Pan, S. L., Yu, J., & Liu, W. (2022). Orchestrating big data analytics capability for sustainability: A study of air pollution management in China. *Information and Management*, *59*(5), 103231. https://doi.org/10.1016/j.im.2019.103231

**APPENDIX**

Table 5: Demographic Profile

| DEMOGRAPHIC PROFILE | CHARACTERISTICS | RESPONDENTS (N=?) | VALID PERCENTAGE (%) |
|---|---|---|---|
| | Total | 303 | 100% |
| GENDER | Male | 210 93 | 69% |
| | Female | | 31% |
| AGE | 31 – 35 Years | 143 | 47.2% |
| | 36 – 40 Years | 48 | 15.8% |
| | 41- 45 Years | 70 | 23.1% |
| | 46 Years and Above | 42 | 13.9% |
| EDUCATION | BBA | 98 | 32.3% |
| | MBA | 101 | 33.3% 10.9% |
| | DIPLOMA OTHER | 33 | 22.4% |
| | | 68 | |
| DEPARTMENT | Marketing | 93 | 30.7% |
| | Operations | 89 | 29.4% |
| | IT | 44 | 14.5% 7.6% |
| | Sales | 23 | 17.8% |
| | Other | 54 | |
| INDUSTRY EXPERIENCE | 1 To 5 Years More | 198 | 65.3% |
| | Than 5 Years | 105 | 34.7% |
| SIZE (COMPANY'S WORKFORCE) | 1-50 Employees | 165 | 54.5% 32.3% |
| | 51-500 Employees | 98 | 13.2% |
| | More Than 500 Employees | 40 | |

*Please indicate your level of agreement or disagreement with the following statements, ranging from 1 = Strongly disagree to 5 = Strongly agree.*

Table 6: Questionnaire

| CONSTRUCT | ITEM | STATEMENTS |
|---|---|---|
| BDA- EDC | SENS | Changes in the internal and external business environment are well tracked by my organization (2019; 2023). |
| | SENS | My firm constantly processes and interprets information (2019; 2023). |
| | SENS | My firm utilizes available opportunities to improve organizational competitiveness (2019; 2023). |
| | SEIZ | My firm uses advanced supporting technologies (2019; 2023). |
| | SEIZ | My firm uses advanced analytical techniques (2019; 2023). |
| | SEIZ | For the purpose of making decisions, my firm integrates and mixes data from several sources (2019; 2023). |
| | RECO | In an unpredictable and complicated environment, my firm often uses data visualization techniques to investigate new market opportunities (2019; 2023). |
| | RECO | In order to reduce risk, my firm employs dashboards for root cause analysis and continuous improvement (2019; 2023). |
| | RECO | My firm provides dashboard apps to our managers' computers and other communication devices (2019; 2023). |
| | RECO | In order to maintain a competitive edge, my firm updates its business strategy largely using data from innovative strategies (2019; 2023). |
| MARKETING AMBIDEXTERITY | EXPLT | To adjust current marketing strategies, my firm regularly reexamines data from previous projects and study (2023; 2011). |
| | EXPLT | My firm routinely adapts current ideas while developing new marketing processes (2023; 2011). |
| | EXPLT | My firm regularly and progressively enhances its current marketing practices (2023; 2011). |
| | EXPLT | To increase efficiency, my firm focuses on making changes to marketing practices (2023; 2011). |
| | MKTP | My firm has demonstrated swift entry into new markets (2023; 2012). |
| MARKET PERFORMANCE | MKTP | My firm has successfully launched new services or products to the marketplace in a timely manner (2023; 2012). |
| | MKTP | My firm has remained highly successful in introducing the recent products or services (2023; 2012). |
| | MKTP | My firm has maintained a strong market share position (2023; 2012). |
| COMPETITOR PRESSURE | COMPRE | Many firms in our industry actively embrace Big Data (2013; 2020). |
| | COMPRE | Big Data Analytics are widely recognized in our industry for providing substantial marketing advantages (2013; 2020). |
| | COMPRE | Within our industry, there is a consensus that leveraging BDA brings significant operational benefits (2013; 2020). |
| | COMPRE | Enhancing organizational image through the adoption of BDA is regarded as crucial in our industry (2013; 2020). |
| | COMPRE | There is a prevailing belief in our industry that the advances derived from BDA outweigh the associated costs (2013; 2020). |
| | COMPRE | The widespread opinion in our industry is that employing BDA is the most appropriate strategy to achieve business objectives (2013; 2020). |

# ABBREVIATION

| | |
|---|---|
| **BDA** | Big Data Analytics |
| **BDA-EDC** | Big Data Analytics Enabled Dynamic Capabilities |
| **BDAC** | Big Data Analytics Capabilities |
| | |
| **SENS** | Sensing |
| **SEIZ** | Seizing |
| **RECO** | Recognition |
| | |
| **EXPLT** | Exploitation |
| | |
| **MKTP** | Market Performance |
| | |
| **COMPRE** | Competitor Pressure |
| | |
| **PLS-SEM** | Partial Least Structural Equation Model |